# ASYMPTOTIC STRUCTURE INCORPORATING DOUBLE AND TRIPLE DECKS, ORIGINAL VERSION

TAREK M. A. EL-MISTIKAWY

Dept. Eng. Math. Phys., Faculty of Engineering, Cairo University, Giza 12211, Egypt.

**ABSTRACT**

The stagnation point flow toward a rotating disk of finite radius is known to develop- near the edge of the disk- into a triple deck structure when the flow is strictly a stagnation point flow, and into a double deck structure when the flow is strictly a rotating disk flow. It is shown here that the transfer from one structure to another is singular; requiring matching of two main decks. The double deck structure is, in essence, a triple deck structure with collapsing upper deck.

## 1. INTRODUCTION

Multi-structured flows are encountered when a high Reynolds number flow experiences rapid streamwise changes. Triple deck structures occur in boundary layers accompanied with streaming outer flow, while double deck structures occur in flows accompanied with quiescent outer flow. Their applications cover several triggering agents: geometric irregularity, mass transfer, incipient shock, local heating, trailing edge …, etc. References to such applications, prior to the year 2000, can be found in the books of Sychev et al. [1] and Sobey [2]. Some recent applications are found in References [3-5].

The analysis of multi-structured flows is based on perturbation techniques. The leading order problem added considerably to our understanding of flow phenomena such as upstream influence, boundary layer separation, and nonlinear instability [2]. The second order problem was derived by Brown and Williams [6], and used by Ragab and Nayfeh [7].

Several attempts aiming at better understanding multi-structured flows; in particular, triple decks; were reported. Meyer [8] established the three notions on which the triple deck theory could be built. Different perturbation techniques [9-11], other than the traditional matched asymptotic technique, were adopted to derive the triple deck structure.

The aim of the present study is to relate the double deck and triple deck structures. It addresses the question whether the double deck structure can be obtained as a limit of the triple deck structure, when the outer flow speed diminishes. A problem that permits such a study is the stagnation point flow toward a rotating disk of finite radius.

The stagnation point flow toward a rotating disk of infinite extent was formulated by Hannah [12]. It incorporates as limiting cases: Homann's [13] axisymmetric stagnation point flow, when the disk is stationary; and von Kármán's [14] rotating disk flow, when the farfield fluid is essentially stationary.

For a disk of finite radius, the way Hannah's flow- which is valid over the disk- develops near the edge of the disk is under consideration. The problem involves two non-dimensional parameters: the Reynolds number Re, and the ratio $\Lambda$ of the strength of the farfield in-flow to the angular speed of the disk. As Re grows indefinitely, a triple deck structure [15] similar to that discovered by Stewartson [16] and Misseter [17] forms in Homann's case ($\Lambda^{-1} = 0$), while in von Kármán's case ($\Lambda = 0$) the double deck structure discovered by Smith [18] and Shidlovskii [19] forms. In Hannah's case, it is shown here that a triple deck structure forms. The lower and upper decks are valid for fixed $\Lambda$ ($0 \leq \Lambda \leq \infty$; the equality to $\infty$ meaning $\Lambda^{-1} = 0$). The upper deck collapses in size and experiences diminishing perturbations when $\Lambda$ becomes $O(\text{Re}^{-1/14})$. The main deck splits into two; corresponding to $\Lambda = O(\text{Re}^0)$ and $\Lambda = O(\text{Re}^{-1/14})$. These two main decks cannot be obtained from one another in a regular manner; i.e. through taking limits. They can, however, be matched.

## 2. FORMULATION OF THE PROBLEM

A steady laminar incompressible stagnation point flow is symmetrically directed toward both sides of a circular disk of radius $L^*$ that is rotating in its plane with uniform angular speed $\omega^*$ about its axis of symmetry. The upper half space is shown in

Fig. 1. Making use of the rotational symmetry, we formulate the problem for a fixed meridional plane on one side of the disk only, and introduce the $z^*$-axis normal to the disk and the $r^*$-axis in the radial direction. The velocity components: $u^*$ in the $r^*$-direction, $v^*$ in the azimuthal direction, and $w^*$ in the $z^*$-direction, as well as the pressure $p^*$ are dependent on $r^*$ and $z^*$ only. Far from the disk, as $z^* \sim \infty$, the flow tends to its inviscid solution with $u^* \sim a^* r^*$, where $a^*$ is a measure of the strength of the stagnation point flow.

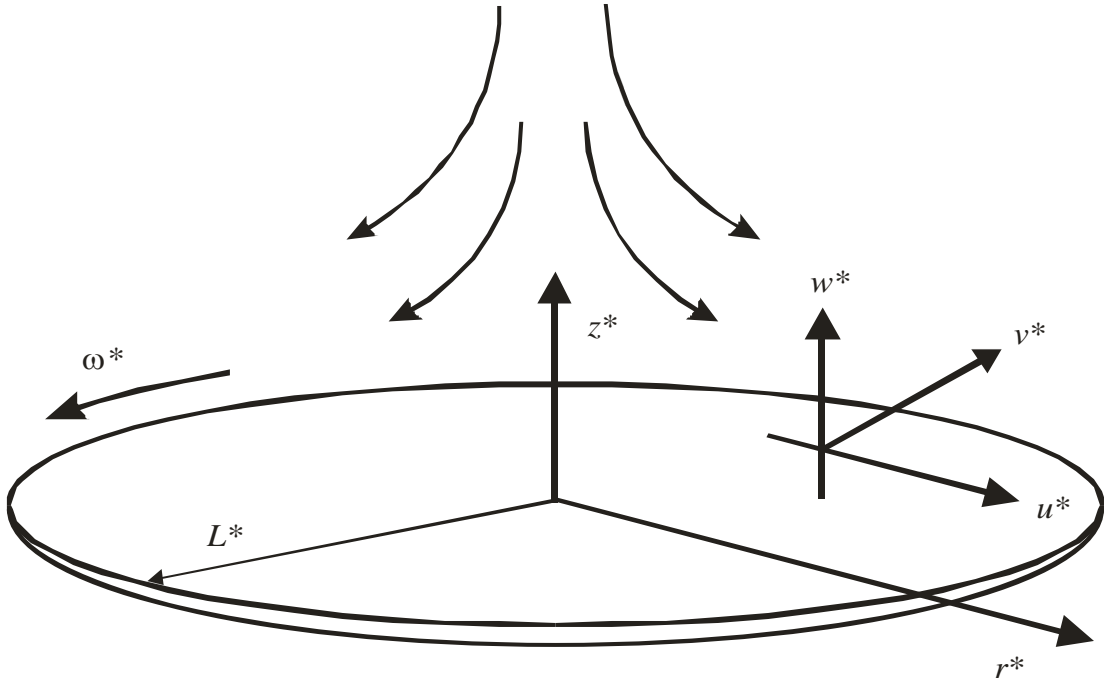

Fig. 1. Flow Configuration.

Non-dimensional variables: $(r, z) = (r^*, z^*)/L^*$, $(u, v, w) = (u^*, v^*, w^*)/U^*$, and $p = (p^* - p_o^*)/\rho^* U^{*2}$ are introduced, where $U^* = L^*(a^{*2} + \omega^{*2})^{1/2}$, $\rho^*$ is the density, and $p_o^*$ is the inviscid stagnation pressure. Also defined are the non-dimensional parameters: the Reynolds number $\text{Re} = \rho^* U^* L^*/\mu^*$, that is taken to be growing indefinitely, and the ratio parameter $\Lambda = a^*/\omega^*$.

The equations governing the flow are the continuity and Navier-Stokes equations:

$$u_{,r} + \frac{u}{r} + w_{,z} = 0$$

$$uu_{,r} - \frac{v^2}{r} + wu_{,z} = \frac{1}{\text{Re}}\left[u_{,rr} + \left(\frac{u}{r}\right)_{,r} + u_{,zz}\right] - p_{,r}$$

$$uv_{,r} + \frac{uv}{r} + wv_{,z} = \frac{1}{\text{Re}}\left[v_{,rr} + \left(\frac{v}{r}\right)_{,r} + v_{,zz}\right]$$

$$uw_{,r} + ww_{,z} = \frac{1}{\text{Re}}\left[w_{,rr} + \frac{1}{r}w_{,r} + w_{,zz}\right] - p_{,z}$$

where, as a convention, subscripts following a comma denote differentiation.

The following boundary conditions apply:

As $z \sim \infty$: the flow tends to the corresponding potential flow [12], so that

$$u \sim \frac{\Lambda}{\sqrt{1+\Lambda^2}} r, \; v \sim 0, \; w \sim -\frac{2\Lambda}{\sqrt{1+\Lambda^2}} z, \; p \sim -\frac{1}{2}\frac{\Lambda^2}{1+\Lambda^2}(r^2 + 4z^2)$$

At $z = 0$, $0 \leq r < 1$: we apply the adherence conditions

$$u = 0, \; v = \frac{1}{\sqrt{1+\Lambda^2}} r, \; w = 0$$

At $z = 0$, $r > 1$: symmetry requires

$$u_{,z} = 0, \; v_{,z} = 0, \; w = 0$$

Introducing the new variables

$$\zeta = \text{Re}^{1/2} z, \; F = u/r, \; G = v/r, \; H = \text{Re}^{1/2} w, \; P = p + \frac{1}{2}\frac{\Lambda^2}{1+\Lambda^2}(r^2 + 4z^2)$$

the governing equations and boundary conditions become

$$rF_{,r} + 2F + H_{,\zeta} = 0$$

$$rFF_{,r} + F^2 - G^2 + HF_{,\zeta} - F_{,\zeta\zeta} = \frac{\Lambda^2}{1+\Lambda^2} - \frac{1}{r}P_{,r} + \frac{1}{\text{Re}}\left[F_{,rr} + \frac{3}{r}F_{,r}\right]$$

$$rFG_{,r} + 2FG + HG_{,\zeta} - G_{,\zeta\zeta} = \frac{1}{\text{Re}}\left[G_{,rr} + \frac{3}{r}G_{,r}\right]$$

$$rFH_{,r} + HH_{,\zeta} - H_{,\zeta\zeta} = 4\frac{\Lambda^2}{1+\Lambda^2}\zeta - \text{Re}P_{,\zeta} + \frac{1}{\text{Re}}\left[H_{,rr} + \frac{1}{r}H_{,r}\right]$$

$$\zeta \sim \infty: F \sim \frac{\Lambda}{\sqrt{1+\Lambda^2}}, G \sim 0, H \sim -\frac{2\Lambda}{\sqrt{1+\Lambda^2}}\zeta, P \sim 0$$

$$\zeta = 0, 0 \leq r < 1: F = 0, G = \frac{1}{\sqrt{1+\Lambda^2}}, H = 0$$

$$\zeta = 0, r > 1: F_{,\zeta} = 0, G_{,\zeta} = 0, H = 0$$

To accommodate the sudden change in the surface conditions at $r = 1$, the flow develops a multi-structure whose suitable perturbation parameter is $\varepsilon = O(\text{Re}^{-1/8})$ in Homann's limit, and $\varepsilon = O(\text{Re}^{-1/7})$ in von Kármán's limit, as Re $\sim\infty$. In either limit, the multi-structure has radial extent $r - 1 = O(\varepsilon^3)$, with main and lower decks characterized by $\zeta = O(\varepsilon^0)$ and $\zeta = O(\varepsilon)$, respectively. These two parameters can be obtained from a new perturbation parameter $\varepsilon \sim 0$ defined by $\text{Re} = \varepsilon^{-8}\Lambda^2$, with $\Lambda = O(\varepsilon^0)$ towards Homann's limit and $\Lambda = O(\varepsilon^{1/2})$ towards von Kármán's limit.

## 3. ASYMPTOTIC ANALYSIS

We carry out a triple deck perturbation analysis, which is- by now- routine. The resulting expansions will, therefore, be presented without much elaboration. Their correctness can be proved by direct substitution in the appropriate equations, and matching the pertinent expansions.

### 3.1 Surface Region { $r < 1, \zeta$ fixed}

Here, the velocity components, being independent of $r$, are expressed as follows:

$$F = \bar{F}(\zeta; \Lambda), G = \bar{G}(\zeta; \Lambda), H = \bar{H}(\zeta; \Lambda)$$

They are governed by the following problem:

$$2\bar{F} + \bar{H}_{,\zeta} = 0$$

$$\bar{F}^2 - \bar{G}^2 + \bar{H}\bar{F}_{,\zeta} - \bar{F}_{,\zeta\zeta} = \frac{\Lambda^2}{1+\Lambda^2}$$

$$2\bar{F}\bar{G} + \bar{H}\bar{\bar{G}}_{,\zeta} - \bar{G}_{,\zeta\zeta} = 0$$

$$\zeta \sim \infty: \bar{F} \sim \frac{\Lambda}{\sqrt{1+\Lambda^2}}, \bar{G} \sim 0$$

$$\zeta = 0: \bar{F} = 0, \bar{G} = \frac{1}{\sqrt{1+\Lambda^2}}, \bar{H} = 0$$

This problem is valid for $0 \leq \Lambda \leq \infty$. Its solution reduces to Homann's when $\Lambda^{-1} = 0$, and approaches von Kármán's when $\Lambda = O(\varepsilon^{1/2}) = \varepsilon^{1/2}\lambda$, as follows

$$\bar{F} \sim \bar{F}_0(\zeta) + \varepsilon^{1/2}\lambda \bar{F}_1(\zeta)$$

$$\bar{G} \sim \bar{G}_0(\zeta) + \varepsilon^{1/2}\lambda \bar{G}_1(\zeta)$$

$$\bar{H} \sim \bar{H}_0(\zeta) + \varepsilon^{1/2}\lambda \bar{H}_1(\zeta)$$

where $\bar{F}_0$, $\bar{G}_0$ and $\bar{\bar{H}}_0$ are von Kármán's velocity components.

In the lower part of this region (i.e., for small $\zeta$), $\bar{F}$, $\bar{G}$ and $\bar{H}$ behave as follows

$$\bar{F} \sim \tau\zeta - \frac{1}{2}\zeta^2 - \frac{1}{3\sqrt{1+\Lambda^2}}\gamma\zeta^3$$

$$\bar{G} - \frac{1}{\sqrt{1+\Lambda^2}} \sim \gamma\zeta + \frac{1}{3\sqrt{1+\Lambda^2}}\tau\zeta^3$$

$$\bar{H} \sim -\tau\zeta^2 + \frac{1}{3}\zeta^3 + \frac{1}{6\sqrt{1+\Lambda^2}}\gamma\zeta^4$$

where $\tau = \bar{F}_{,\zeta}(0;\Lambda)$ and $\gamma = \bar{G}_{,\zeta}(0;\Lambda)$ represent the radial and azimuthal shear stresses, respectively.

### 3.2 Near-Wake Region $\{r \downarrow 1\}$

Taking $r = 1 + \tau\xi^3$ and letting $\xi \sim 0$, this region splits into a main layer in which $\zeta = O(\xi^0)$, and a lower layer in which $\zeta = O(\xi) = \xi\eta$. The following matching expansions for the velocity components are obtained.

Main layer expansions:

$$F - \bar{F} \sim -\alpha_1 \bar{F}_{,\zeta}\xi$$

$$G - \bar{G} \sim -\alpha_1 \bar{G}_{,\zeta}\xi$$

$$H \sim \frac{\alpha_1}{3\tau\xi^2}\bar{F}$$

Lower layer expansions:

$$F \sim \xi\tau\phi_1'$$

$$G - \frac{1}{\sqrt{1+\Lambda^2}} \sim \xi\gamma\phi_1'$$

$$H \sim \frac{1}{3\xi}(\eta\phi_1' - 2\phi_1)$$

where $\phi_1(\eta)$ and $\alpha_1$ satisfy

$$\phi_1''' + \frac{2}{3}\phi_1\phi_1'' - \frac{1}{3}\phi_1'^2 = 0$$

with

$$\phi_1(0) = 0,\ \phi_1''(0) = 0,\ \phi_1''(\infty) = 1,\ \alpha_1 = \lim_{\eta\to\infty}(\eta - \phi_1')$$

with a dash denoting differentiation with respect to $\eta$.

### 3.3 Edge Region $\{r = 1 + \varepsilon^3 x;\ x \text{ fixed}\}$

This region joins the surface region to the near-wake region. It is comprised of three decks: an upper deck $\{\zeta = O(\varepsilon^{-1}\Lambda)\}$, a main deck $\{\zeta = O(\varepsilon^0)\}$, and a lower deck $\{\zeta = O(\varepsilon)\}$.

### 3.3.1 Upper deck $\{\zeta = \varepsilon^{-1}\Lambda N;\ N \text{ fixed}\}$

With $H$ expressed as

$$H \sim \varepsilon^{-2} H_1(x, N; \Lambda) + \varepsilon^{-1} H_2(x, N; \Lambda)$$

the flow variables expand as follows

$$F - \frac{\Lambda}{\sqrt{1+\Lambda^2}} \sim \Lambda^{-1}[\varepsilon^2 \frac{1}{\pi} \overline{\int_{-\infty}^{\infty} \frac{(x-X)H_1(X,0;\Lambda)}{(x-X)^2 + N^2} dX}$$

$$+ \varepsilon^3 \frac{1}{\pi} \overline{\int_{-\infty}^{\infty} \frac{(x-X)H_2(X,0;\Lambda)}{(x-X)^2 + N^2} dX}]$$

$$G \sim 0$$

$$H + \frac{2\Lambda}{\sqrt{1+\Lambda^2}} \zeta \sim \varepsilon^{-2} \frac{1}{\pi} \overline{\int_{-\infty}^{\infty} \frac{NH_1(X,0;\Lambda)}{(x-X)^2 + N^2} dX} + \varepsilon^{-1} \frac{1}{\pi} \overline{\int_{-\infty}^{\infty} \frac{NH_2(X,0;\Lambda)}{(x-X)^2 + N^2} dX}]$$

$$P \sim -\frac{1}{\sqrt{1+\Lambda^2}} [\varepsilon^2 \frac{1}{\pi} \overline{\int_{-\infty}^{\infty} \frac{(x-X)H_1(X,0;\Lambda)}{(x-X)^2 + N^2} dX} + \varepsilon^3 \frac{1}{\pi} \overline{\int_{-\infty}^{\infty} \frac{(x-X)H_2(X,0;\Lambda)}{(x-X)^2 + N^2} dX}]$$

where the over-bars denote Cauchy's principal value of the integrals.

**3.3.2 Lower deck** { $\zeta = \varepsilon n$; $n$ fixed }

The flow variables expand as follows

$$F \sim \varepsilon f_1 + \varepsilon^2 f_2$$
$$G - \frac{1}{\sqrt{1+\Lambda^2}} \sim \varepsilon g_1 + \varepsilon^2 g_2$$
$$H \sim \varepsilon^{-1} h_1 + h_2$$
$$P \sim \varepsilon^2 q_1(x) + \varepsilon^3 q_2(x)$$

The ( )$_1$-variables satisfy the following governing equations and boundary conditions:

$$f_{1,x} + h_{1,n} = 0 \tag{1}$$

$$f_1 f_{1,x} + h_1 f_{1,n} - f_{1,nn} = -q_{1,x} \tag{2}$$

$$f_1 g_{1,x} + h_1 g_{1,n} - g_{1nn} = 0 \tag{3}$$

$$n = 0,\ x < 0:\ f_1 = 0,\ g_1 = 0,\ h_1 = 0 \tag{4-6}$$

$$n = 0,\ x > 0:\ f_{1,n} = 0,\ g_{1,n} = 0,\ h_1 = 0 \tag{7-9}$$

$$x \sim -\infty:\ f_1 \sim \tau n,\ g_1 \sim \gamma n,\ q_1 \sim 0 \tag{10-12}$$

$$x \sim \infty:\ f_1 \sim \tau(n - \alpha_1 \tau^{-1/3} x^{1/3}),\ g_1 \sim \gamma(n - \alpha_1 \tau^{-1/3} x^{1/3}),\ q_1 \sim 0 \tag{13-15}$$

Matching conditions to the main deck remain to be invoked.

### 3.3.3 Main deck { $\zeta$ fixed}

As Hannah's flow advances along the main deck, it experiences a lateral shift

$$\varepsilon D \sim \varepsilon \overline{D}_1(x; \Lambda) + \varepsilon^2 [\overline{D}_2(x; \Lambda) + \widehat{D}_2(x, \zeta; \Lambda)]$$

where

$$\varepsilon \overline{D} \sim \varepsilon \overline{D}_1 + \varepsilon^2 \overline{D}_2$$

is a laterally uniform displacement caused by the viscous lower deck, while

$$\varepsilon^2 \widehat{D}_2 = \varepsilon^2 Q_1 \int^{\zeta} \frac{1}{\overline{F}^2} d\zeta = \varepsilon^2 Q_1 \int^{\zeta} (\frac{1}{\overline{F}^2} - \frac{1}{\tau^2 \zeta^2} + \frac{1}{\tau^2 \zeta^2}) d\zeta$$

$$= \varepsilon^2 Q_1 \left[ \int_0^{\zeta} (\frac{1}{\overline{F}^2} - \frac{1}{\tau^2 \zeta^2}) d\zeta - \frac{1}{\tau^2 \zeta} \right]$$

represents the stream-tube divergence in the main deck due to the pressure perturbation $P \sim \varepsilon^2 Q_1(x; \Lambda)$.

To proceed further, we have to consider cases of fixed $\Lambda$ and fixed $\lambda$, separately.

### 3.3.3.1 Main deck I { $\Lambda$ fixed}

Here, the flow variables expand as follows

$$F \sim \bar{F} - \varepsilon \bar{D}_1 \bar{F}_{,\zeta} + \varepsilon^2 \left[ \tfrac{1}{2}\bar{D}_1^2 \bar{F}_{,\zeta\zeta} - \left( \bar{D}_2 + Q_1 \int^\zeta \frac{1}{\bar{F}^2} d\zeta \right) \bar{F}_{,\zeta} - \frac{Q_1}{\bar{F}} \right]$$

$$+ \varepsilon^3 \left[ \Lambda^{-2} \bar{D}_{1,xx} \left\{ \bar{F}_{,\zeta} \int_0^\zeta \left( \frac{1}{\bar{F}^2} \int_0^\zeta \bar{F}^2 d\zeta \right) d\zeta + \frac{1}{\bar{F}} \int_0^\zeta \bar{F}^2 d\zeta \right\} + \cdots \right]$$

$$G \sim \bar{G} - \varepsilon \bar{D}_1 \bar{G}_{,\zeta} + \varepsilon^2 \left[ \tfrac{1}{2}\bar{D}_1^2 \bar{G}_{,\zeta\zeta} - \left( \bar{D}_2 + Q_1 \int^\zeta \frac{1}{\bar{F}^2} d\zeta \right) \bar{G}_{,\zeta} \right]$$

$$+ \varepsilon^3 \left[ \Lambda^{-2} \bar{D}_{1,xx} \bar{G}_{,\zeta} \int_0^\zeta \left( \frac{1}{\bar{F}^2} \int_0^\zeta \bar{F}^2 d\zeta \right) d\zeta + \cdots \right]$$

$$H \sim \varepsilon^{-2} \bar{D}_{1,x} \bar{F} + \varepsilon^{-1} \left[ -\bar{D}_1 \bar{D}_{1,x} \bar{F}_{,\zeta} + \left( \bar{D}_{2,x} + Q_{1,x} \int^\zeta \frac{1}{\bar{F}^2} d\zeta \right) \bar{F} \right]$$

$$- \left[ \Lambda^{-2} \bar{D}_{1,xxx} \bar{F} \int_0^\zeta \left( \frac{1}{\bar{F}^2} \int_0^\zeta \bar{F}^2 d\zeta \right) d\zeta + \cdots \right]$$

$$P \sim \varepsilon^2 Q_1 + \varepsilon^3 \left[ Q_2(x; \Lambda) - \Lambda^{-2} \bar{D}_{1,xx} \int_0^\zeta \bar{F}^2 d\zeta \right]$$

Obviously, the expansions behave in a singular manner as $\Lambda \sim 0$, with the $O(\varepsilon^3)$-terms becoming of the same order as the $O(\varepsilon^2)$-terms when $\Lambda = O(\varepsilon^{1/2})$. A conjugate main deck is, thus, needed to resolve this singularity.

### 3.3.3.2 Main deck II { $\lambda$ fixed}

Here, we have the following expansions:

$$F \sim \bar{F}_0 + \varepsilon^{1/2} \lambda \bar{F}_1 + \varepsilon (\lambda^2 \bar{F}_2 - \bar{D}_{10} \bar{F}_{0,\zeta}) + \varepsilon^{\tfrac{3}{2}} [\lambda^3 \bar{F}_3 - \lambda (\bar{D}_{10} \bar{F}_{1,\zeta} + \bar{D}_{11} \bar{F}_{0,\zeta})]$$

$$+ \varepsilon^2 \Big[ \lambda^4 \bar{F}_4 - \lambda^2 (\bar{D}_{10} \bar{F}_{2,\zeta} + \bar{D}_{11} \bar{F}_{1,\zeta} + \bar{D}_{12} \bar{F}_{0,\zeta})$$

$$- \left( \bar{D}_{20} + Q_{10} \int^\zeta \frac{1}{\bar{F}_0^{\,2}} d\zeta \right) \bar{F}_{0,\zeta} + \tfrac{1}{2} \bar{D}_{10}^2 \bar{F}_{0,\zeta\zeta} - \frac{Q_{10}}{\bar{F}_0}$$

$$+ \lambda^{-2} \bar{D}_{10,xx} \left\{ \bar{F}_{0,\zeta} \int_0^\zeta \left( \frac{1}{\bar{F}_0^{\,2}} \int_0^\zeta \bar{F}_0^{\,2} d\zeta \right) d\zeta + \frac{1}{\bar{F}_0} \int_0^\zeta \bar{F}_0^{\,2} d\zeta \right\} \Big]$$

$$G \sim \bar{G}_0 + \varepsilon^{1/2}\lambda\bar{G}_1 + \varepsilon(\lambda^2\bar{G}_2 - \bar{D}_{10}\bar{G}_{0,\zeta}) + \varepsilon^{\frac{3}{2}}[\lambda^3\bar{G}_3 - \lambda(\bar{D}_{10}\bar{G}_{1,\zeta} + \bar{D}_{11}\bar{G}_{0,\zeta})]$$
$$+ \varepsilon^2\left[\lambda^4\bar{G}_4 - \lambda^2(\bar{D}_{10}\bar{G}_{2,\zeta} + \bar{D}_{11}\bar{G}_{1,\zeta} + \bar{D}_{12}\bar{G}_{0,\zeta})\right.$$
$$-\left(\bar{D}_{20} + Q_{10}\int^{\zeta}\frac{1}{\bar{F}_0^{\,2}}d\zeta\right)\bar{G}_{0,\zeta} + \tfrac{1}{2}\bar{D}_{10}^2\bar{G}_{0,\zeta\zeta}$$
$$\left. + \lambda^{-2}\bar{D}_{10,xx}\bar{G}_{0,\zeta}\int_0^{\zeta}\left(\frac{1}{\bar{F}_0^{\,2}}\int_0^{\zeta}\bar{F}_0^{\,2}d\zeta\right)d\zeta\right]$$

$$H \sim \varepsilon^{-2}\bar{D}_{10,x}\bar{F}_0 + \varepsilon^{-\frac{3}{2}}\lambda(\bar{D}_{10,x}\bar{F}_1 + \bar{D}_{11,x}\bar{F}_0)$$
$$+ \varepsilon^{-1}\left[\lambda^2(\bar{D}_{10,x}\bar{F}_2 + \bar{D}_{11,x}\bar{F}_1 + \bar{D}_{12,x}\bar{F}_0) + \left(\bar{D}_{20,x} + Q_{10,x}\int^{\zeta}\frac{1}{\bar{F}_0^{\,2}}d\zeta\right)\bar{F}_0\right.$$
$$\left. - \bar{D}_{10}\bar{D}_{10,x}\bar{F}_{0,\zeta} - \lambda^{-2}\bar{D}_{10,xxx}\bar{F}_0\int_0^{\zeta}\left(\frac{1}{\bar{F}_0^{\,2}}\int_0^{\zeta}\bar{F}_0^{\,2}d\zeta\right)d\zeta\right]$$

$$P \sim \varepsilon^2[Q_{10} - \lambda^{-2}\bar{D}_{10,xx}\int_0^{\zeta}\bar{F}_0^{\,2}d\zeta]$$

These expansions match correctly to those of the Main Deck I.

### 3.3.4 Closure

The following additive composite expressions for the two main decks can be formed

$$F \sim \bar{F} - \varepsilon\bar{D}_1\bar{F}_{,\zeta} + \varepsilon^2\left[\tfrac{1}{2}\bar{D}_1^2\bar{F}_{,\zeta\zeta} - \left(\bar{D}_2 + Q_1\int^{\zeta}\frac{1}{\bar{F}^2}d\zeta\right)\bar{F}_{,\zeta} - \frac{Q_1}{\bar{F}}\right]$$

$$G \sim \bar{G} - \varepsilon\bar{D}_1\bar{G}_{,\zeta} + \varepsilon^2\left[\tfrac{1}{2}\bar{D}_1^2\bar{G}_{,\zeta\zeta} - \left(\bar{D}_2 + Q_1\int^{\zeta}\frac{1}{\bar{F}^2}d\zeta\right)\bar{G}_{,\zeta}\right]$$

$$H \sim \varepsilon^{-2}\bar{D}_{1,x}\bar{F} + \varepsilon^{-1}\left[-\bar{D}_1\bar{D}_{1,x}\bar{F}_{,\zeta} + \left(\bar{D}_{2,x} + Q_{1,x}\int^{\zeta}\frac{1}{\bar{F}^2}d\zeta\right)\bar{F}\right]$$

$$P \sim \varepsilon^2[Q_1 - \lambda^{-2}\bar{D}_{10,xx}\int_0^{\zeta}\bar{F}_0^{\,2}d\zeta]$$

Matching to the lower deck gives

$$Q_1 = q_1$$

as well as the following conditions on the lower deck problem

$$n \sim \infty: \lim_{\varepsilon \to 0}[f_1 - \tau(n - \bar{D}_1)] = 0, \; \lim_{\varepsilon \to 0}\left[g_1 - \gamma(n - \bar{D}_1) + \frac{\gamma q_1}{\tau^2 n}\right] = 0 \quad (16,17)$$

Matching to the upper deck, on the other hand, gives

$$H_1(X, 0; \Lambda) = \frac{\Lambda}{\sqrt{1 + \Lambda^2}} \bar{D}_{1,x}$$

and, consequently,

$$\lim_{\varepsilon \to 0}\left[q_1 - \lambda^{-2}\bar{D}_{10,xx} \int_0^\infty \bar{F}_0^{\,2} d\zeta - \frac{\Lambda}{1 + \Lambda^2} \frac{1}{\pi} \int_{-\infty}^\infty \frac{\overline{\bar{D}_{1,x}}}{(x - X)} dX\right] = 0 \quad (18)$$

This is the interaction law relating the lower deck pressure $q_1$ to the displacement function $\bar{D}_1$. For fixed $\Lambda$,

$$q_1 = -\frac{\Lambda}{1 + \Lambda^2} \frac{1}{\pi} \int_{-\infty}^\infty \frac{\overline{\bar{D}_{1,x}}}{(x - X)} dX$$

whereas, for fixed $\lambda$,

$$q_1 = \lambda^{-2}\bar{D}_{10,xx} \int_0^\infty \bar{F}_0^{\,2} d\zeta$$

The leading order lower deck problem is now complete; governed by Eqs. (1-18).

## 4. CONCLUSION

By considering the stagnation point flow toward a rotating disk of finite radius, it has been shown that the double deck and triple deck structures can be incorporated in a single matching asymptotic structure. A new perturbation parameter $\varepsilon$; incorporating the two non-dimensional parameters of the problem: the Reynolds number Re and the ratio parameter $\Lambda$, has been identified. For $\varepsilon \sim 0$, the triple deck structure shown in Fig. 2, has been established in the three-dimensional space $(x, \zeta; \Lambda)$. In its interaction region, the structure involves a lower deck and an upper deck, both extending for $0 < \Lambda < \infty$, and two neighboring main decks, for $\Lambda = O(\varepsilon^0)$ and $\Lambda = O(\varepsilon^2)$. The upper deck perturbations diminish, in a regular manner, as $\Lambda \sim 0$; and the deck itself collapses in size, ultimately vanishing. The structure, thus, reduces to a double deck structure.

The case $\lambda = 0$ (The case $\Lambda = \infty$) can be obtained in a regular manner from that of fixed $\lambda$ (fixed $\Lambda$); by invoking the transformations $\varepsilon = \delta \acute{\varepsilon}$, $x = \delta^{-3}\acute{x}$, $N = \delta^{-3}\acute{N}$, $\overline{D}_1 = \delta^{-1}\acute{\overline{D}}_1$, $P_1 = \delta^{-2}\acute{P}_1$, $n = \delta^{-1}\acute{n}$, $f_1 = \delta^{-1}\acute{f}_1$, $g_1 = \delta^{-1}\acute{g}_1$, $h_1 = \delta \acute{h}_1$, $q_1 = \delta^{-2}\acute{q}_1$, ..., where $\delta = \lambda^{2/7}$ ($\delta = \Lambda^{1/4}$), and then setting $\lambda = 0$ ($\Lambda = \infty$).

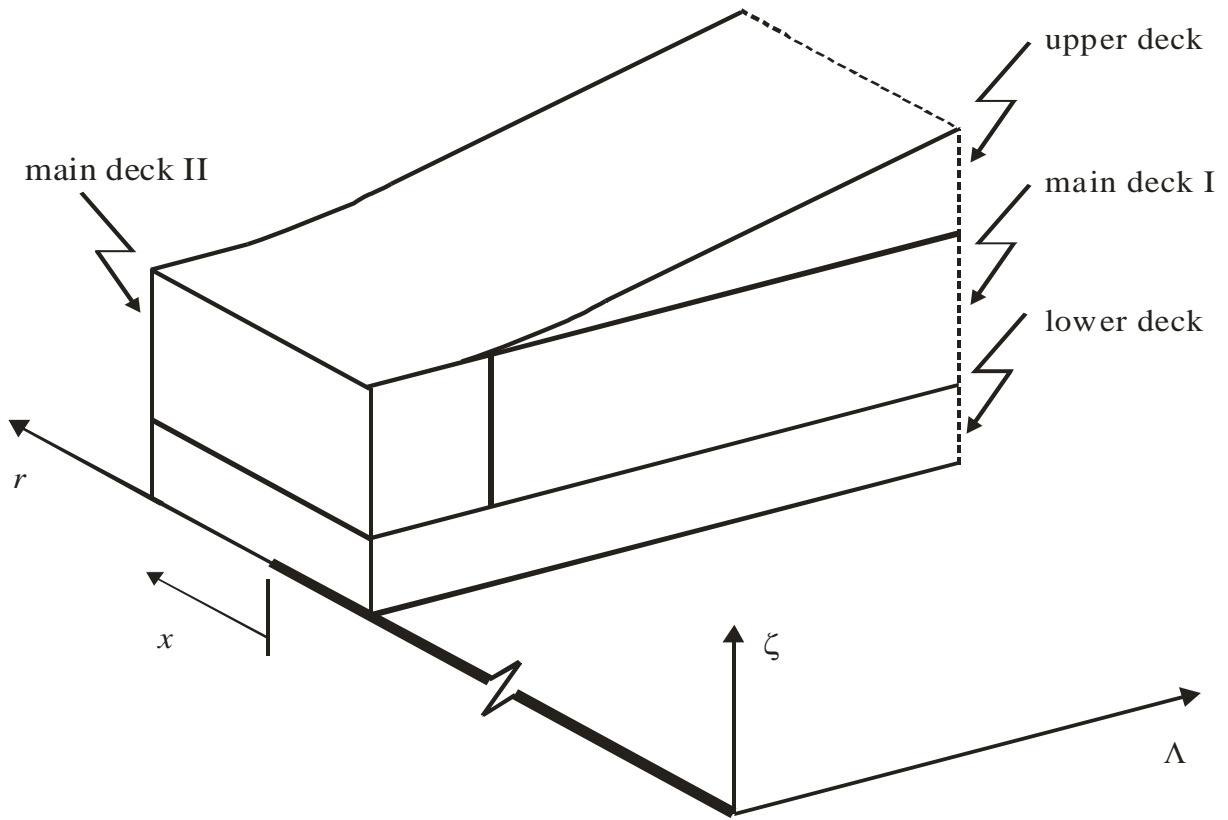

Fig. 2. Asymptotic Structure

**Note:** The intriguing problem of the stagnation point flow toward a finite rotating disk motivated on-and-off study over a span of 20 years. How to relate two asymptotic structures posing as infinitely distant boundaries in the $(r, z, \Lambda)$ space, the double deck structure (at $\Lambda = 0$) and the triple deck structure (at $\Lambda = \infty$)? When submitted for publication, the problem was criticized for dealing with an impractical configuration. The analysis was readily applied to the simpler case of the flat plate [20]; where only the double deck structure posed as a boundary (at $\Lambda = 0$) to the $(x, y, \Lambda)$ space.